\documentclass[journal]{IEEEtran}
\usepackage{epsfig}
\usepackage{graphicx}
\usepackage{caption}
\usepackage{subcaption}
\usepackage{balance}
\usepackage{float}
\usepackage{psfrag}
\usepackage{amsfonts,amsmath,amssymb,mathrsfs,txfonts}

\usepackage{url}
\usepackage{cite}
\usepackage{verbatim}
\usepackage{multirow}
\usepackage{multicol}
\usepackage[font=small]{caption}
\usepackage{pifont}
\usepackage{algorithm,algpseudocode}
\usepackage{xcolor}
\begin{document}
\title{ Firefly Algorithm for Movable Antenna Arrays}
\author{Manh Kha Hoang, {\it Member IEEE}, Tuan Anh Le, {\it Senior Member IEEE}, Kieu-Xuan Thuc, {\it Member IEEE}, Tong Van Luyen, {\it Member IEEE}, Xin-She Yang, and Derrick Wing Kwan Ng, {\it Fellow IEEE} 
\thanks{M. K. Hoang, T. V. Luyen, and K.-X. Thuc are with the Faculty of Electronics Engineering, Hanoi University of Industry, Hanoi, Vietnam. Email: \{khahoang; luyentv; thuckx\}@haui.edu.vn.
}
\thanks{T. A. Le and X.-S. Yang are with the Department of Design Engineering \& Mathematics, Faculty of Science and Technology, Middlesex University, The Burroughs, Hendon, London, NW4 4BT, U. K. Email: \{t.le; x.yang\}@mdx.ac.uk.}
\thanks{D. W. K. Ng is with the University of New South Wales, Sydney, NSW, Australia. Email: w.k.ng@unsw.edu.au.}
}
\markboth{IEEE Wireless Communications Letters, DOI: 10.1109/LWC.2024.3456899 }%
{Shell \MakeLowercase{\textit{et al.}}: Bare Demo of IEEEtran.cls for IEEE Journals}
\maketitle

\begin{abstract}
This letter addresses a multivariate optimization problem for linear movable antenna arrays (MAAs). Particularly, the position and beamforming vectors of the under-investigated MAA are optimized simultaneously to maximize the minimum beamforming gain  across several intended directions, while ensuring interference levels at various unintended directions remain below specified thresholds. To this end, a swarm-intelligence-based firefly algorithm (FA) is introduced to acquire an effective solution to the optimization problem. Simulation results reveal the superior performance of the proposed FA approach compared to the state-of-the-art approach employing alternating optimization and successive convex approximation. This is attributed to the FA's effectiveness in handling non-convex multivariate and multimodal optimization problems without resorting approximations.
\end{abstract}
\IEEEpeerreviewmaketitle
\section{Introduction}

The problems of coupling variables in joint beamforming and position design for movable antenna arrays (MAAs) are typically tackled by utilizing the sub-optimal alternating optimization (AO) approach \cite{MARuizhang2024,Yang10458417,Zhang10474119,Chen10437006,Mei10508218,Hu10414081,Hu10416363,Feng10504625,Lai10375698}. Specifically, this method decomposes the original problem into two sub-problems, where each of the optimization variables is solved in an iterative manner while keeping the other fixed, e.g., \cite{MARuizhang2024,Hu10416363,Zhang10474119,Feng10504625,Hu10414081}. Unfortunately, these decomposed sub-problems are mostly non-convex, hence, require further manipulations to acquire an effective solution. For example, the successive convex approximation (SCA) technique was employed in \cite{MARuizhang2024} to establish convex surrogate problem counterparts. Similarly, SCA was leveraged in \cite{Chen10437006,Hu10414081}, and \cite{Feng10504625} for tackling the sub-problem for the position vector. 

Despite the fruitful efforts devoted in the literature, the solutions to beamforming and position vectors in MAA problems, derived by the exploitation of AO and other approximation methods, cannot be guaranteed as globally optimal. More importantly, the performance gaps between these suboptimal solutions and their corresponding globally optimal counterparts have never been revealed. Motivated by this fact, this letter employs a nature-inspired meta-heuristic firefly algorithm (FA), which has been shown to be more effective in solving non-convex beamforming problems than the interior point method \cite{Tuan10311527}, to find an effective solution for a MAA problem. 

This letter jointly optimizes the beamforming and position vectors of a linear MAA such that the minimum beamforming gain across various intended directions is maximized subject to maintaining interference levels at multiple unintended directions below required thresholds. This multivariate optimization problem is known to be highly non-convexity \cite{MARuizhang2024,Hu10414081}. Unlike the state-of-the-art approach in \cite{MARuizhang2024}, which addresses independently yet iteratively the beamforming and position vectors in each approximated convex sub-problem, this letter proposes a novel swarm-intelligence-based FA to concurrently optimize both vectors without any approximation. To this end, the penalty method \cite{YangFA2008} is first exploited to convert the originally non-convexity problem to an equivalent unconstrained problem and then linearly mapping it to the flashing brightness of a firefly which does not alter the originality of the problem. A firefy population is then initiated in which each firefly is associated with a pair of randomly generated position and beamforming vectors.  The nature behaviour of the tropical fireflies, i.e., flying toward the brighter one with some random moves along their ways, will finally lead them to the brightest one representing the solution to the problem. Furthermore, the computational complexity of the proposed approach is thoroughly analyzed. Simulation results confirm the convergence of the propose FA approach as well as its superior performance over the state-of-the-art approach in \cite{MARuizhang2024}. This letter extends our recent work in \cite{Tuan10311527}, originally introduced for transmit beamforming designs, to a receive beamforming problem in MAAs. To the best of the authors' knowledge, this is the first work leveraging a nature-inspired FA approach in MAAs.

\emph{\textbf{Notation}:}
Lower and upper case letters $s$ and $S$: a scalar; bold-lower-case letter $\mathbf{s}$: a column vector; $(\cdot)^T$: the transpose operator; $(\cdot)^H$: the complex-conjugate-transpose operator; $\left\|\cdot\right\|$: the Euclidean norm;  $\mathbb{C}^{N_A\times 1}$: the set of $N_A\times 1$ complex-element vectors; $\mathbb{R}^{N_A\times 1}$: the set of $N_A\times 1$ real-element vectors.

\section{Problem Formulation}

Consider a linear antenna array of $N_A$ antennas and each of which is movable, i.e., its location can be relocated, within an one-dimensional line segment\footnote{The adopted approach in this letter can be easily extended to two-dimensional or three-dimensional MAAs by representing the MAA elements' positions in a matrix form.} of length $L$ \cite{MARuizhang2024,Zhang10474119,Hu10414081,Hu10416363}. Let $d_i\in [0,L]$ be the location of the $i$-th antenna, $\forall i\in\{1,2,\cdots,N_A\}$, i.e., $0\leq d_{1}\leq d_2\leq \cdots \leq d_{N_A}\leq L$,  and $\mathbf{d}=[d_1, d_2, \cdots,d_{N_A}]^T \in \mathbb{R}^{N_A\times 1}$. Let $\theta$ be the steering angle with respect to the line/axis joining all elements of the array and $\psi$ be the carrier wavelength, the antenna steering vector $\mathbf{s}\left( \mathbf{d},\theta\right)\in \mathbb{C}^{N_A \times 1}$ for a receive MAA can be expressed as\footnote{ Since the focus of this letter is to design position and beamforming vectors to amplify the impinging signals at intended directions while suppressing the interference at unintended angles, the channel effects, e.g., path-loss and fading, are not considered, as also adopted in e.g., \cite{MARuizhang2024,Hu10416363}. }:
\begin{equation}
   \mathbf{s}\left( \mathbf{d},\theta\right)=
    \left[ e^{j\frac{2 \pi}{\psi}d_1\cos{\theta}}, e^{j\frac{2 \pi}{\psi}d_2\cos{\theta}},\cdots,e^{j\frac{2 \pi}{\psi}d_{N_A}\cos{\theta}}\right]^T.
\end{equation}
Let $\mathbf{w}\in \mathbb{C}^{N_A \times 1}$ be the beamforming vector, the beamforming gain with respect to angle $\theta$ can be written as:
\begin{eqnarray}
    G\left( \mathbf{w}, \mathbf{d},\theta\right)=|\mathbf{w}^H\mathbf{s}\left( \mathbf{d},\theta\right)|^2.
\end{eqnarray}

Without loss of generality, let $\{ \theta_t\}^T_{t=1}$ and $\{ \phi_q\}^Q_{q=1}$ be the set of $T$ desired signal directions and the set of $Q$ undesired interference directions, respectively. We aim to jointly determine the beamforming vector $\mathbf{w}$ and the location vector $\mathbf{d}$ such that the minimum beamforming gain is maximized while keeping the  power at undesired directions below the required threshold, $I_0$. To this end, the optimization problem can be posed as:
\begin{equation} \label{MA01}
\begin{aligned}
& \underset{\{\mathbf{w} \}, \{\mathbf{d} \}} {\textrm{max}} \ \underset{ t }{\textrm{min}} & &
\left\{G\left( \mathbf{w}, \mathbf{d},\theta_t\right)\right\}\\
& \mbox{s.\ t.}\ & & C1: d_1\geq 0, \\
&&& C2: d_{N_A}\leq L,\\
&&& C3:d_i - d_{i-1} \geq L_0, \ \forall i=2,3,\cdots, N_A,\\
&&& C4: G\left( \mathbf{w}, \mathbf{d},\phi_q\right)\leq I_0, \forall q=1,2,\cdots, Q,\\
&&& C5: ||\mathbf{w}||_2 \leq 1,
\end{aligned}
\end{equation}
where $L_0$ is the minimum distance between two adjacent antenna elements. Besides, $\{\mathbf{w} \}$ and $ \{\mathbf{d} \}$ are two sets of different variables to be optimized. 
$C1$, $C2$, and $C3$ are imposed to ensure that the antennas are within the range $[0,L]$ and the distance between a pair of neighbouring antennas is always not smaller than a required value of $L_0$ to avoid coupling effect \cite{MARuizhang2024}. $C4$ is to control the interference level at the undesired directions below the threshold. Finally, $C5$ is to guarantee the normalized power of the beamforming vector.

In general, problem\footnote{Problems \eqref{MA01} in this letter and (P1) in \cite{MARuizhang2024} are equivalent as (P1) is an epigraph form of \eqref{MA01} where the non-linear objective in \eqref{MA01} is transformed into a constraint in (P1) by introducing an auxiliary optimization variable $\delta$. } \eqref{MA01} is highly non-convex due to two reasons: i) its objective and constraint $C4$ are non-concave with respect to either  $\mathbf{w}$ or $\mathbf{d}$; ii) the two sets of optimization variables, $\mathbf{w}$ and $\mathbf{d}$, are coupled in the objective function and $C4$  \cite{Hu10414081}. In the following, we  introduce the FA to tackle the problem.

\section{Proposed FA Approach}
To facilitate the presentation, let $f_i=L_0-d_i +d_{i-1}$ and $\Phi_q\left(\phi_q\right)=|\mathbf{w}^H\mathbf{s}\left( \mathbf{d},\phi_q\right)|^2-I_0$.  Then, we can rewrite \eqref{MA01} as:
\begin{equation} \label{MAFA}
\begin{aligned}
& \underset{\{\mathbf{w} \},\{ \mathbf{d}\} } {\textrm{max}} \ \underset{ t }{\textrm{min}} & &
\left\{G\left( \mathbf{w}, \mathbf{d},\theta_t\right)\right\}\\
& \mbox{s.\ t.}\ & & -d_1 \leq 0,\\
&&& d_{N_A}-L \leq 0,\\
&&& f_i \leq 0,\forall i,i=2,3,\cdots, N_A,\\
&&&\Phi_q\left( \phi_q\right) \leq 0, \forall q=1,2,3,\cdots, Q, \\
&&& ||\mathbf{w}||_2-1\leq 0.
\end{aligned}
\end{equation}

We then proceed by utilizing the penalty method \cite{YangFA2008} to transform problem \eqref{MAFA} into an equivalent un-constrained problem:
\begin{equation} \label{IRS_PerfectCSI_n2}
\begin{aligned}
\underset{\{\mathbf{w} \}, \{\mathbf{d} \}} {\textrm{max}} & &
\left\{G\left( \mathbf{w}, \mathbf{d}\right)+ P\left(\mathbf{w},\mathbf{d},\theta_t,\phi_q\right)\right\},
\end{aligned}
\end{equation}
where $ G\left( \mathbf{w}, \mathbf{d}\right)=\underset{ t }{\textrm{min}}  \left\{
G\left( \mathbf{w}, \mathbf{d},\theta_t\right)\right\}$
and $P\left(\mathbf{w},\mathbf{d},\theta_t,\phi_q\right)$ is the penalty term given as:
\begin{eqnarray}
P\left(\mathbf{w},\mathbf{d},\theta_t,\phi_q\right)&=&\beta_1\text{max}\left\{0,-d_1 \right\}^2+\beta_2\text{max}\left\{0,d_{N_A}-L \right\}^2\nonumber\\
   &+&\sum_{i=2}^{N_A}\beta_{3,i}\text{max}\left\{0,f_i\right\}^2\nonumber\\&+&\sum_{q=1}^{Q}\rho_q \text{max}\left\{0, \Phi_q(\phi_q)\right\}^2\nonumber\\
   &+&\lambda \text{max}\left\{0, ||\mathbf{w}||_2-1\right\}^2,\label{penalty}
\end{eqnarray}
with $\beta_1>0$, $\beta_2>0$, $\beta_{3,i}>0$, $\lambda>0$ and $\rho_q>0$ are penalty constants. 

Note that the FA is developed based on the following three idealized rules \cite{YangFA2008,YangFA2009}. First, fireflies are assumed to be unisex and attract others within their population. Second, the attractiveness of any firefly to others is proportional to its flashing brightness. Both attractiveness and flashing brightness decrease as the distance between two fireflies increases. Given any two flashing fireflies, the less brighter firefly will fly towards the brighter mate. If a firefly does not find any brighter one, it will randomly fly. Third, the flashing brightness of a firefly depends on the landscape of the objective function. Their relations with solution search in the considered problem will be discussed later.

In this letter, we adopt the generalized FA approach \cite{Tuan10311527} to solve the receive beamforming problem in \eqref{MAFA} with two sets of independent optimization variables $\{\mathbf{w}\}$ and $\{\mathbf{d}\}$. To start with, let $\left\{\mathbf{w}_k,\mathbf{d}_k\right\}$ be firefly $k$. Specifically, we randomly initialize a population of $\Omega$ fireflies $\{\mathbf{w}_k,\mathbf{d}_k\}$, $k\in \{1,2,\cdots, \Omega\}$, and the third rule is implemented by defining the flashing brightness, i.e., the light density,  of firefly $k$  as: 

\begin{equation}
   B_k\left(\mathbf{w}_k,\mathbf{d}_k\right)=
G\left( \mathbf{w}_k, \mathbf{d}_k\right)+ P\left(\mathbf{w}_k,\mathbf{d}_k,\theta_t,\phi_q\right). \label{lightRIS}
\end{equation}

On the other hand, the first and second rules are interpreted by considering any fireflies $k$ and $j$ amongst the population, if $B_k\left(\mathbf{w}_k,\mathbf{d}_k\right) > B_j\left(\mathbf{w}_j,\mathbf{d}_j\right)$, then  firefly $j$ will fly toward firefly $k$ at the $n$-th generation as:
\begin{eqnarray}
    \mathbf{w}_j^{(n+1)}&=&\mathbf{w}_j^{(n)}+\beta_0 e^{- \left(\gamma^{0.5}r_{w,kj}^{(n)}\right)^2}\left(\mathbf{w}_k^{(n)}-\mathbf{w}_j^{(n)} \right)+\alpha^{(n)}\mathbf{\tilde{w}},\label{FAmoveW}\\
    \mathbf{d}_j^{(n+1)}&=&\mathbf{d}_j^{(n)}+\beta_0 e^{- \left(\gamma^{0.5}r_{d,kj}^{(n)}\right)^2}\left(\mathbf{d}_k^{(n)}-\mathbf{d}_j^{(n)} \right)+\alpha^{(n)}\mathbf{\tilde{d}},\label{FAmoveTheta}
\end{eqnarray}
where $r_{w,kj}^{(n)}=|| (\mathbf{w}_k^{(n)}-\mathbf{w}_j^{(n)}||$ and $r_{d,kj}^{(n)}=|| (\mathbf{d}_k^{(n)}-\mathbf{d}_j^{(n)}||$ are the Cartesian distances,  $\beta_0$ is the attractiveness at $r_{w,kj}^{(n)}=0$ and $r_{d,kj}^{(n)}=0$, $\gamma$ presents the variation of the attractiveness. Note that the second terms of equations \eqref{FAmoveW} and \eqref{FAmoveTheta} capture the attractions while the third terms of these equations  are  randomizations comprised of a randomization factor $\alpha^{(n)}\in[0,1]$ associated with two random vectors $\mathbf{\tilde{w}} \in\mathbb{C}^{N_A \times 1} $ and   $\mathbf{\tilde{d}} \in\mathbb{R}^{N_A \times 1}$. The real and imaginary parts of each element of $\mathbf{\tilde{w}}$ and the entries of $\mathbf{\tilde{d}} $ are drawn from either an uniform or a Gaussian distribution. The proposed FA for solving optimization problem \eqref{MAFA} is described in Algorithm~\ref{FA_MAA}.

The value $\gamma^{-0.5}$ determines the average distance of a flock of fireflies which is seen by its neighbour flocks. Therefore, the whole population is automatically divided into different flocks enabling the ability of solving highly non-linear and multimodal problems \cite{Tuan10311527}. The randomizations in \eqref{FAmoveW} and \eqref{FAmoveTheta} are in fact random walks in which the exploitation and exploration of the algorithm can be managed by controlling the value of $\alpha^{(n)}$. When the population size significantly exceeds  the number of local optima, the iterations of Algorithm~1 will stochastically guide the initial population toward the best solution amongst the local optima. Indeed, the population can theoretically attain the global optimal solution if $\Omega\rightarrow \infty$ and $n\gg 1$ \cite{YangFA2008,YangFA2009,Tuan10311527}. In fact, it has been reported in \cite{Tuan10311527} that the FA converges with less than 120 generations for four representative transmit beamforming approaches with fixed-position antenna array scenarios.

\begin{algorithm}
	\caption{Firefly Algorithm for Solving \eqref{MAFA}}
	\label{FA_MAA}
	\begin{algorithmic}[1]
		\State \textbf{Input:} Two sets $\{ \theta_k\}^T_{t=1}$ and $\{ \phi_q\}^Q_{q=1}$; $L$; $N_A$; $L_0$;  $I_0$; $\Omega$; maximum generation $R$; $\beta_1$; $\beta_2$; $\beta_{3,i}$; $\lambda$; $\rho_q$; $\beta_0$; $\gamma$;
		\State Randomly generate $\Omega$ populations $\left\{\{\mathbf{w}_1,\mathbf{d}_1\}, \{\mathbf{w}_2,\mathbf{d}_2\},\cdots, \{\mathbf{w}_{\Omega},\mathbf{d}_{\Omega}\}\right\}$;
		\State Calculate the flashing brightness of $\Omega$ fireflies as \eqref{lightRIS};
		\State  Sort the firefly population in a descending order of the flashing brightness $B_k\left(\mathbf{w}_k,\mathbf{d}_k\right)$;
		\State Declare the current best solution: $B^{\star}:=B_1\left(\mathbf{w}^{\star},\mathbf{d}^{\star}\right)$; $\{\mathbf{w}^{\star},\mathbf{d}^{\star}\}:=\{\mathbf{w}_1,\mathbf{d}_1\}$;
		\For{$n =1:R$}
		    \For{$j=1:\Omega$}
		    \For{$k=1:\Omega$}
		    \If{$B_j\left(\mathbf{w}_j,\mathbf{d}_j\right)>B^{\star}$} 
		    \State $B^{\star}:=B_j\left(\mathbf{w}_j,\mathbf{d}_j\right)$; $\{\mathbf{w}^{\star},\mathbf{d}^{\star}\}:=\{\mathbf{w}_j,\mathbf{d}_j\}$;
		    \EndIf
		    \If{$B_k\left(\mathbf{w}_k,\mathbf{d}_k\right)>B^{\star}$} 
		    \State $B^{\star}:=B_k\left(\mathbf{w}_k,\mathbf{d}_k\right)$; $\{\mathbf{w}^{\star},\mathbf{d}^{\star}\}:=\{\mathbf{w}_k,\mathbf{d}_k\}$;
		    \EndIf
		    \If{$B_k\left(\mathbf{w}_k,\mathbf{d}_k\right)>B_j\left(\mathbf{w}_j,\mathbf{d}_j\right)$}
		\State Move firefly $j$ towards firefly $k$ as \eqref{FAmoveW}, \eqref{FAmoveTheta};
		    \EndIf
		\State Attractiveness varies with distances via $e^{- \left(\gamma^{0.5}r_{w,kj}^{(n)}\right)^2}$ and $e^{-\left(\gamma^{0.5}r_{d,kj}^{(n)}\right)^2}$;
		\State Evaluate new solutions and update light intensity as \eqref{lightRIS};
		    \EndFor 
		    \EndFor 
		    \State Sort the firefly population in a descending order  of the flashing brightness $B_k\left(\mathbf{w}_k,\mathbf{d}_k\right)$;
		    \State Update the current best solution: $B^{\star}:=B_1\left(\mathbf{w}^{\star},\mathbf{d}^{\star}\right)$; $\{\mathbf{w}^{\star},\mathbf{d}^{\star}\}:=\{\mathbf{w}_1,\mathbf{d}_1\}$;
		\EndFor
		\State \Return $\mathbf{w}^{\star}, \mathbf{d}^{\star}$.
		
	\end{algorithmic}
\end{algorithm}

{\it Computational complexity:} It is clear that the computational complexity of the proposed FA approach presented in Algorithm~\ref{FA_MAA} is dominated by the three loops from steps 6 to 24. With a note that $xyz$ is the order of computational complexity of evaluating the product of two matrices of sizes $x \times y$ and $y \times z$ adopting a schoolbook iterative algorithm \cite{Tuan10311527}, one can verify that the computational complexities of steps 16, 18, and 19 are on the order of $N_A$. Moreover the computational complexity of step 22 is on the order of $\Omega \log{\Omega}$. Therefore, one can conclude that the dominant term of the computational complexity of Algorithm~\ref{FA_MAA} is on the order of $R \left(\Omega^2 N_A+ \Omega \log{\Omega}\right)$.

\section{Simulation Results}
\subsection{Simulation Setup}
The array length is set at $L=8\psi$ with the minimum distance between two adjacent antennas of $L_0=\psi/2$. The setup parameters for FA are as follows. The variation of the attractiveness $\gamma$ is set as 1. The penalty constants are set identical but they dynamically vary as $\beta_1=\beta_2=\beta_{3,i}=\lambda=\rho_q= n^2, \ \forall i\in\{2,3,\cdots,N_A\}$, where $n$ is the generation index in Algorithm~\ref{FA_MAA}. The attractiveness at zero distance is $\beta_0=1$ \cite{Tuan10311527}. Finally, the initial randomization factor is $\alpha^{(0)}=0.07$  and its value at the $n$-th generation is $\alpha^{(n)}=\alpha^{(0)}0.989^n$.
\subsection{Performance Comparison }\label{FA_vs_AO_SCA}
Here, the performance of the proposed FA approach is evaluated and compared with the state-of-the-art approach in \cite{MARuizhang2024}, hereafter is referred to as AO-SCA.  
Monte Carlo simulations are carried out over 50 direction distributions where in each distribution, a number of $T$ intended directions and $Q$ unintended directions are randomly generated following an uniform distribution between $[0^{\circ},180^{\circ}]$ with the step size of $5^{\circ}$. Furthermore, for each direction distribution, Algorithm~\ref{FA_MAA} is executed 50 times to obtain an average performance. The firefly population and number of generations are, respectively, $\Omega=40$ and $R=500$. 
\begin{figure}[t]
\centering
    \includegraphics[width=.4\textwidth]{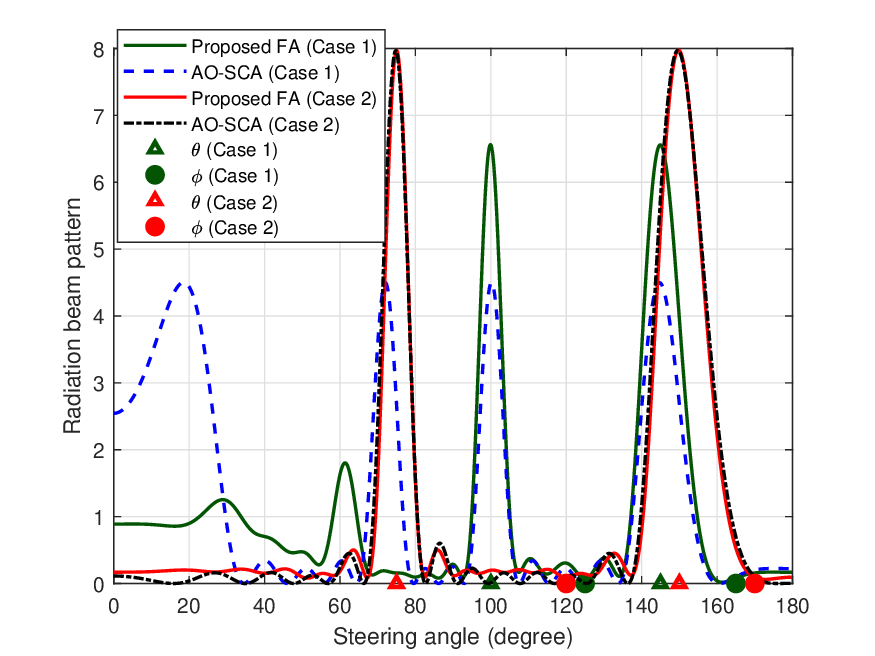}
\caption{Radiation beam patterns. $N_A=8$. $L_0=\frac{\psi}{2}$; $L=8\psi$. $T=2$; $Q=2$; and $I_0=0.1$. Case 1: $\{ \theta_t\}^T_{t=1}=[100^{\circ}, 145^{\circ}]$ and $\{ \phi_q\}^Q_{q=1}=[125^{\circ}, 165^{\circ}]$. Case 2: $\{ \theta_t\}^T_{t=1}=[75^{\circ}, 150^{\circ}]$ and $\{ \phi_q\}^Q_{q=1}=[120^{\circ}, 170^{\circ}]$. }\label{gain8an}
\end{figure}

In Fig.~\ref{gain8an}, the radiation beamforming patterns of the proposed FA and the AO-SCA are shown for two following cases of direction distributions. In Case 1, two intended directions are located on the same side of the boresight of the antenna array, i.e., $\{ \theta_t\}^2_{t=1}=[100^{\circ}, 145^{\circ}]$. In Case 2, two intended directions are located on different sides of the boresight of the antenna array, i.e., $\{ \theta_t\}^2_{t=1}=[75^{\circ}, 150^{\circ}]$. It can be seen from the figure that both proposed FA and the AO-SCA approaches attain almost $100 \%$ beamforming gain for Case 2. However, for Case 1, the proposed FA outperforms the AO-SCA with the max-min beamforming gain of $6.56$, i.e., $82\%$ beamforming gain, in comparison with  $4.48$, i.e., $56\%$ beamforming gain, offered by the AO-SCA. The same trends as seen in Case 1 and Case 2 were also observed in other direction distributions. 

These above results can be explained as follows. Under favorable conditions, i.e., Case 2 distributions, the surrogate functions developed in the SCA in \cite{MARuizhang2024} do not deviate from these original ones. As a result, the AO-SCA can attain almost the highest possible gain. The approximation-free FA approach, as expected, also offers similar performance as its counterpart. On the other hand, under unfavorable conditions, i.e., Case 1 distributions, the AO-SCA suffers from a high side-lobe problem, i.e., the side lobes have unexpected high beamforming gains, indicating a fact that the approximations imposed by the AO and SCA processes severely degrade the system performance as they are unable to effectively control the potential energy leakage. In fact, the AO-SCA approach is highly likely trapped in local optimums resulting in sub-optimal solutions \cite{MARuizhang2024,Hu10416363}. The proposed FA, on the other hand, is free from any kind of approximations and has a well control on its side lobes, hence, offers a much better solution than its counterpart. This confirms the effectiveness of the proposed FA in handling multivariate and non-convex problems. 

\begin{figure}[t]
\centering
    \includegraphics[width=.4\textwidth]{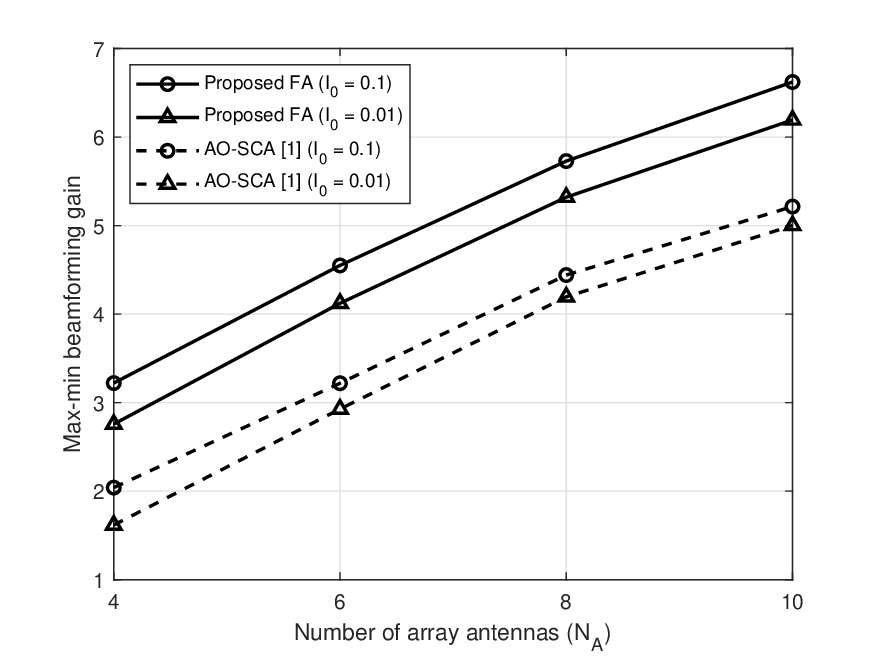}
\caption{ Max-min beamforming gain versus the number of array antennas $N_A$ when $T=Q=2$. }\label{gainvsNt}
\end{figure}

Fig.~\ref{gainvsNt} illustrates the max-min beamforming gain versus the number of array antenna elements $N_A$ for the proposed FA and AO-SCA approaches with  different values of the interference threshold $I_0$. It is clear that the proposed FA prevails the AO-SCA for all setups of $I_0$ and $N_A$. For instance, the former offers the max-min beamforming gain higher than the latter around 1.2 and 1.3 for $I_0=0.1$ and $I_0=0.01$, respectively.
\begin{figure}[t]
\centering
    \includegraphics[width=.4\textwidth]{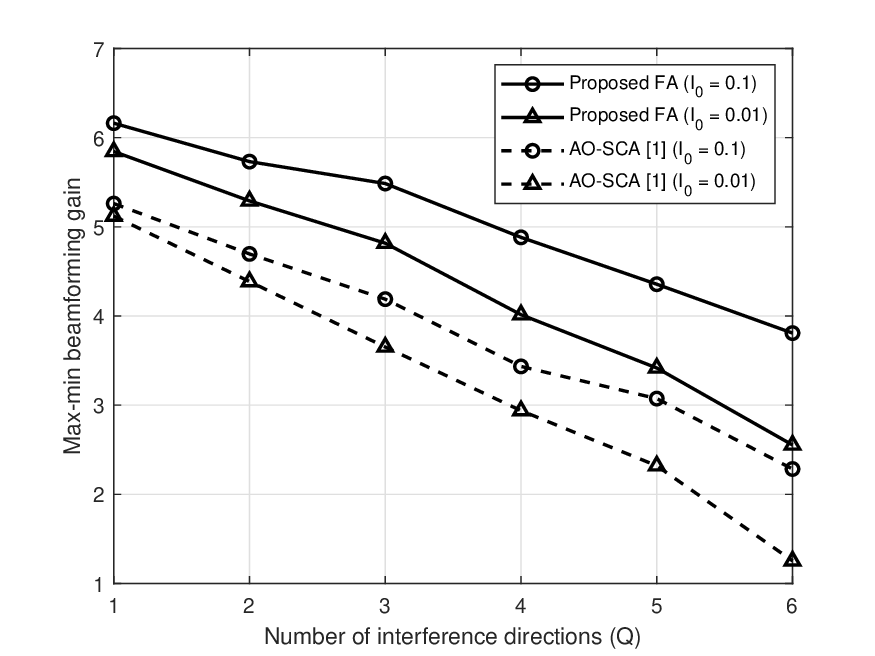}
\caption{ Max-min beamforming gain versus the number of undesired interference directions $Q$ when $T=2$, $N_A=8$.}\label{gainvsQ}
\end{figure}

Fig.~\ref{gainvsQ} indicates the max-min beamforming gain versus the number of interference directions with different values of $I_0$. Fig.~\ref{gainvsQ} again confirms the superior performance of the proposed FA over the AO-SCA. Interestingly, the proposed FA are capable of handling tougher scenarios with higher number of interference directions while the AO-SCA struggles to do so. This can be observed by the fact that the performance gap between the proposed approach and its counterpart enlarges when the number of interference directions increases. For example, the gaps are, respectively, 0.90 and 1.53 at $Q=1$ and $Q=6$ with $I_0=0.1$. When $I_0=0.01$, the gaps are, respectively, 0.72 and 1.30. The less effective behaviour of the AO-SCA is again due to the approximation process. The superior performance of the proposed approach is due to its approximation-free nature as well as the exploitation and exploration capabilities of the FA.

\subsection{Impacts of $\Omega$ and R}
In this section, the performance of the proposed FA approach is evaluated via Monte Carlo simulations with 50 executions of Algorithm~\ref{FA_MAA} for one direction distribution  with $\{ \theta_t\}^T_{t=1}=[100^{\circ}, 145^{\circ}]$ and $\{ \phi_q\}^Q_{q=1}=[125^{\circ}, 165^{\circ}]$, i.e., Case 1.
\begin{figure}[t]
\centering
    \includegraphics[width=.4\textwidth]{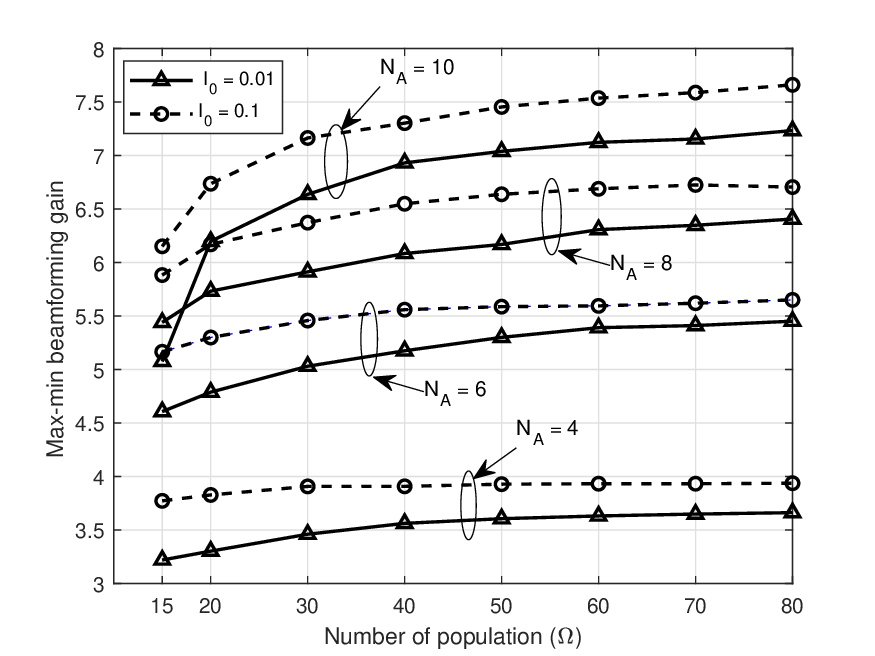}
\caption{ Max-min beamforming gain versus maximum number of population $\Omega$ with different numbers of $N_A$ and $I_0$ when $T=Q=2$.}\label{gainvsOmega}
\end{figure}

Fig.~\ref{gainvsOmega} plots the max-min beamforming gain versus the number of firefly population $\Omega$ and different values of $I_0$ and $N_A$. The results reveal the fact that the beamforming gain significantly improves when the number of population increases from 15 to 40, except for the case of $N_A=4$ and $I_0=0.1$. For example, with $N_A=10$ when the population grows from 15 to 40, the gain improves from 6.15 to 7.30 and from 5.08 to 6.93 for $I_0=0.1$ and $I_0=0.01$, respectively. These results are due to the fact that a larger population size provides a better representation of the feasibility region hence allowing the FA to reach a better solution. However, a population of around 20 fireflies is sufficient to reflect the global optimality for the setup $\{N_A=4,\ I_0=0.1\}$. Therefore, having more fireflies does not result in noticeable improvement. This trend is also seen with $\{N_A=6,\ I_0=0.1\}$ where 30 fireflies are adequate to solve the problem.

On the other hand, Fig.~\ref{gainvsOmega} also reveals that the gain increases 0.3 with $N_A=8$ and 0.36 with $N_A=10$ if the population size grows from 40 to 80. This is due to the fact that a higher population size is required to represent the enlarger feasible space as a result of increased numbers of antennas.

\begin{figure}[t]
\centering
    \includegraphics[width=.4\textwidth]{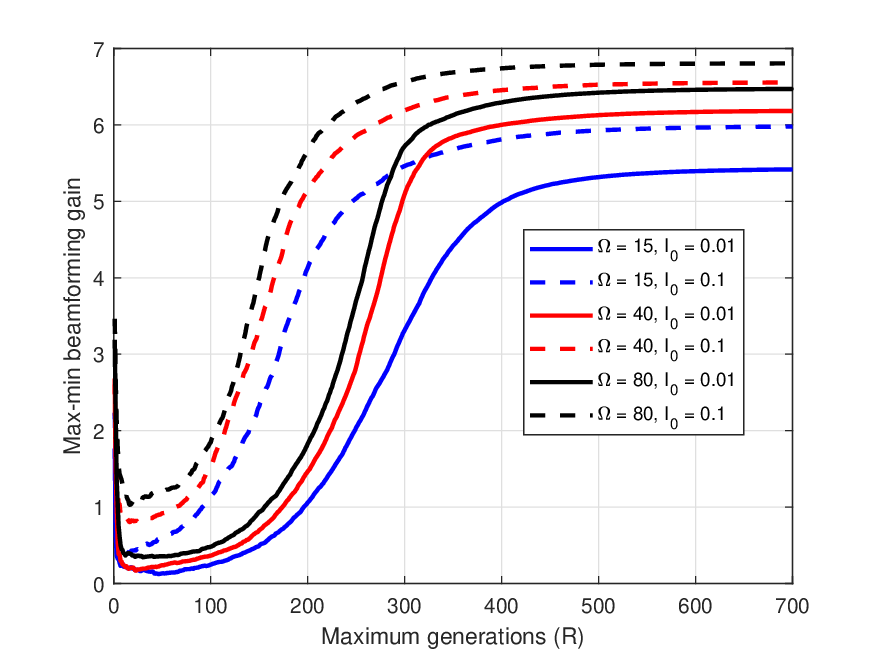}
\caption{ Max-min beamforming gain versus the maximum generations $R$ with different values of $\Omega$ and $I_0$ when $T=Q=2$ and $N_A=8$.}\label{gainvsR}
\end{figure}

Fig.~\ref{gainvsR} shows the max-min beamforming gain versus the  maximum generations $R$ with different population sizes. It is clear from the figure that the beamforming gains converge after around 500 generations for all setups. 
\subsection{CPU Run Time Comparison}
The simulations were carried out on a Matlab R2022b on a 64-bit-Lubuntu-Operating-System server equipped with 40 virtual CPUs and 256 GB RAM. The total processing speed was equivalent to 94.687 GHz. CVX package \cite{Boyd} was employed to implement the AO-SCA approach in \cite{MARuizhang2024}.
 \begin{figure}[t]
\centering
    \includegraphics[width=.4\textwidth]{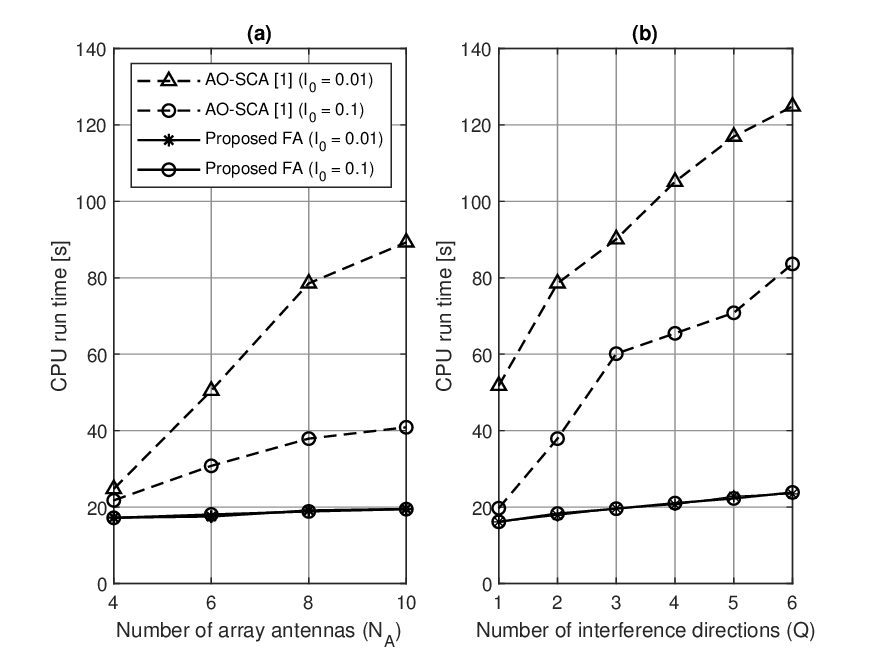}
\caption{CPU run time versus: (a) the number of antennas when $T=Q=2$; (b) the number of interference directions when $T=2$, $N_A=8$.}\label{CPU}
\end{figure}
The CPU run time for the experiments shown in Figs. \ref{gainvsNt} and \ref{gainvsQ} are, respectively, plotted in Figs. \ref{CPU} (a) and (b).

It is clear from Fig. \ref{CPU} that the proposed FA outperforms the AO-SCA approach with a lower CPU run time.  The run-time gap significantly increases with higher values of $N_A$ and $Q$, or lower value of $I_0$. The results can be explained as follows. The AO-SCA involves three types of iterative processes/loops. Two inner loops are dedicated to solving two SCA problems, where either beamforming or position vector is iteratively solved with  initiative values of itself and the other vector in each problem. The outer loop repetitively solves these two SCA problems until convergence. In challenging scenarios, i.e., when $N_A$ and $Q$ increase or $I_0$ decreases, due to these inherited approximations, the AO-SCA suffers from high numbers of inner and outer iterations to obtain a solution leading to a longer CPU run time. On the other hand, the proposed FA, being free from any approximations, maintains a consistent number of calculations resulting in a relatively stable and lowest CPU run time. The computational complexity of the FA does not depend on the value of $I_0$ while it is mainly determined by the number of fireflies $\Omega$ and the maximum generation $R$. Consequently,  as shown on the figure, the values of $N_A$ and $Q$ have a small impact on the CPU run time while the value of $I_0$ does not.
\section{Conclusion}
This letter proposed a novel nature-inspired FA to solve a multivariate, highly non-convex optimization problem in MAAs.  
Simulations confirmed the superior performances, i.e., max-min beamforming gain and CPU run time, of the proposed FA over the state-of-the-art AO-SCA approach.  

\bibliographystyle{IEEEtran}

\end{document}